\documentclass[aip,
 jap,numerical,
 amsmath,amssymb,
 reprint,%
]{revtex4-1}

\usepackage{graphicx}
\usepackage{dcolumn}
\usepackage{bm}

\usepackage[utf8]{inputenc}
\usepackage[T1]{fontenc}
\usepackage{mathptmx}
\usepackage{etoolbox}
\usepackage{bbm}
\usepackage{subfigure}
\usepackage{color}

\makeatletter
\def\@email#1#2{%
 \endgroup
 \patchcmd{\titleblock@produce}
  {\frontmatter@RRAPformat}
  {\frontmatter@RRAPformat{\produce@RRAP{*#1\href{mailto:#2}{#2}}}\frontmatter@RRAPformat}
  {}{}
}%

\makeatother
\begin{document}

\preprint{AIP/123-QED}

\title[Multifractal wave functions of charge carriers in graphene ...]{ Multifractal wave functions  of charge carriers in graphene with folded deformations, ripples or uniaxial flexural modes: analogies to the quantum Hall effect under random pseudomagnetic fields.}
\author{Abdiel Espinosa-Champo and Gerardo G. Naumis}
 \email{naumis@fisica.unam.mx}
\affiliation{ 
Departamento de Sistemas Complejos, Instituto de Fisica, Universidad Nacional Aut\'onoma de M\'exico, Apartado Postal 20-364,01000,Ciudad de M\'exico, M\'exico.
}%

\date{10 September 2021}

\begin{abstract}
The electronic behavior in graphene under arbitrary uniaxial deformations, such as foldings or flexural fields is studied  by including in the Dirac equation pseudoelectromagnetic fields.  General foldings are thus studied by showing that uniaxial deformations can be considered pseudomagnetic fields in the Coulomb gauge norm. This allows to give an expression for the Fermi (zero) energy modes wavefunctions. For random deformations, contact is made
with previous works on the quantum Hall effect under random magnetic fields, showing that the density of states has a power law behavior and that the zero energy modes wavefunctions are multifractal.  This hints at an unusual electron velocity distribution. Also, it is shown that a strong Aharonov-Bohm pseudo-effect is produced. For more general non-uniaxial general flexural strain, it is not possible to use the Coulomb gauge. The results presented here helps to tailor-made graphene uniaxial deformations to achieve specific wave-functions. 
\end{abstract}

\maketitle

\section{Introduction}
Recently, Dirac materials have attracted intense research interest following the celebrated discovery of a two-dimensional (2D) hexagonal allotropic atomic carbon, graphene \cite{Novoselov666}, because of its peculiar band structure and its fascinating properties  \cite{AlessandroCresti2008,LuisEF2014}
largely due to the massless Dirac fermion behavior of the charge carriers.

Due to such excellent mechanical, magnetic and thermal properties of graphite monolayers, they can be used for the development of superconducting devices for micro-electromechanical and nano-electromechanical systems, leading to the development of the next generation of nanoelectronics \cite{RevModPhys.81.109,RevModPhys.83.407}.
As the use of graphene sheets increases, the understanding of the mechanical behaviour is necessary and important for the design and analysis of graphene nanostructures and nanosystems. This opened a new field of research known as straintronics, which aims to refine the electronic and optical properties by applying mechanical deformations  \cite{Naumis2017}. 
Following this direction, many theoretical works have been made studying the effect of mechanical strains on the electronic properties  \cite{GGNaumis2009,Rodriguez2016} using a tight-binding approach  \cite{Zhang2010,Zhang2011} and effective Hamiltonians for low energies  in the vicinity of Dirac points  \cite{Vozmediano2010,Oliva2013,Oliva2015,Guinea2009}. These electronic degrees of freedom are coupled to the structural lattice deformations, and this allows to modify its electronic properties in interesting ways \cite{Vozmediano2010,Amorim2015,Bastos2014, Chen2016, Deji2017}. It has been shown that a model to describe the coupling of the electrons to the out-of-plane deformation should be the Dirac equation in curved space  \cite{Amorim2015,Volovik2014,Volovik2015,RichardKerner2012}. Such coupling is due to the appearance of pseudo-magnetic fields caused by the deformations \cite{Vozmediano2010,Oliva2013,Oliva2015,Naumis2017,Oliva2016,Bastos2018}, 
{\color{black} and leads to a weak localisation/antilocalisation crossover \cite{Falko}.   Mesoscopic conductance fluctuations in graphene  have also been studied by  using diagrammatic
perturbation theory \cite{Falko_Meso}. Yet,  recent experiments with graphene show unexplained exotic multifractal conductance fluctuations  around the Dirac point \cite{Amin2018} (zero modes).}

Moreover, in recent years experimental evidence has been found that for certain regimes, fluctuations in graphene membranes follow a Cauchy distribution that results in large movements and sudden changes in curvature by means of the  \textit{mirror buckling} effect \cite{Ackerman2014, Ackerman2016,Thibado2014}.  This mirror buckling effect was first related to the heating due to the scanning microscope. Later on, it was found that this mirror buckling is always presents and that the height of the flexural vibrations follow a L\'evy distribution with parameters $\alpha=1.5, \gamma=0$ \cite{Ackerman2016}. It was also found an unusual distribution of electron velocities  \cite{Ackerman2016} and a theory has been proposed to explain it  \cite{Kai2019}.  However, this last theory is based on considering carbon atoms in the framework of the classical kinetic theory of gases and the Fokker-Planck-Kolmogorov master equation, but this scheme does not explicitly consider the contribution of  out-of-plane acoustic modes and that the membrane executes Brownian motion with rare large height excursion indicative of L\'evy walks. Thus, a more exhaustive study is needed concerning this point. Likewise, Mao et. al. \cite{Mao2020}  demonstrate that graphene monolayers placed on an atomically flat substrate can be forced to undergo a buckling transition, resulting in a periodically modulated pseudo-magnetic field, which in turn creates a `post-graphene' material with flat electronic bands. This buckling of 2D crystals offers a strategy for exploring interaction phenomena characteristic of flat bands. 

In addition, there is an growing interest in folded deformations  due to  transport properties of strained folds in graphene exhibit a rich behavior ranging from Coulomb blockade to Fabry-P\'erot oscillations for different fold orientations. Those exhibiting strong confinement, behave as electronic waveguides in the direction parallel to the fold axis, providing a new way to realize 1D conducting channels in 2D graphene by strain engineering \cite{Sandler2018}. In general, the mechanical displacements on graphene causes strong changes in the vacuum-induced shifts of the transition frequency of some emitter and, because its low mass and high $Q$ factor, make it a particular attractive candidate for a wide class of sensors \cite{Muschik2014}.

Most previous work concerning this topic has been focused in studying the electron mobility  thorough using transport equations \cite{Castro_2010,Pereira2019}.  In the present work, we study the effects on charge carriers due to the presence of pseudo-electromagnetic fields which models the case of vertical fluctuations due to folded deformations or flexural modes. These modes have a large phonon population originating from the quadratic phonon dispersion and are known to dominate the electron scattering \cite{Castro_2010} and thermal transport \cite{Feng2018,Balandin2020}. In particular, we show that for certain kind of flexural fields, one can make close contact with previous works on Dirac fermions in random electromagnetic potentials, besides its close relationship with the phase transition between the plateaus in Hall's quantum states and the quasi-excitations in \textit{d}-wave superconductors \cite{Ichinose2002}. Then we show that for more general fields, the Coulomb gauge condition used in this work can not be fulfilled.

 It is important to remark that the methods presented here can be extended to study other optoelectronic properties in 2D materials, such as phosphorene 
\cite{Mehboudi5888} or borophene \cite{Naumis2017}, 
and these effects can also be studied using the present methodology, as plane deformations or flexural waves can be considered as random pseudo-electromagnetic waves; in addition, the present results can be extended for new Dirac materials  \cite{doi:10.1093/nsr/nwu080, doi:10.1080/00018732.2014.927109}.

The work is organized as follows. In Sec.  \ref{Model Hamiltonian}, we introduce the effective Hamiltonian for low energies and obtain the time-independent Schrödinger equation to be solved.  In Sec. \ref{sec:Folded deformations}, we analize specifically the electronic properties of graphene with folded deformations. And finally, we present the conclusions in Section \ref{sec: conclusions}.

\section{HAMILTONIAN MODEL \label{Model Hamiltonian}} 
\begin{figure}[h!]
\includegraphics[scale=0.6]{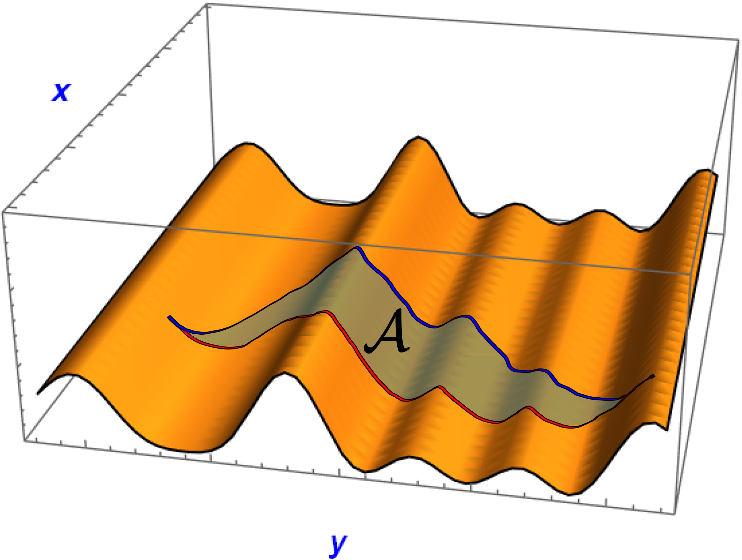}
\caption{{\color{black}Random ripples of a graphene sheet. Two possible electron paths which enclose the area $\mathcal{A}$ are indicated.}
}
\label{fig:Ripple}
\end{figure}

Out-of-plane acoustic modes are characteristic vibrations in graphene. These low frequency modes, seen in Fig. \ref{fig:Ripple}, are easy to excite and carry most of the vibrational energy \cite{Jiang2015,Bastos2018}. They consist in a dynamic elongation, bending and torsion of the local bonds. The stretching or tension of the bonds is by far the most important for the electrons, since it causes a greater impact on the tunneling parameter \cite{RevModPhys.81.109}. Some lattice deformations can be expressed by a gauge field using a Hamiltonian at low energies \cite{Bastos2018, Vozmediano2010}.

{\color{black}The low-energy Hamiltonian for non-interacting electrons in deformed graphene for flexural deformations has been investigated  qualitatively and quantitatively in the literature \cite{Morpurgo_2006,Guinea2009,Rainis_Guinea_2011,Bastos2018,Sasaki2008}. It consists in a Dirac equation added with pseudoelectromagnetic effective fields plus additional contributions caused by several mechanisms, as for example, a $\pi-\sigma$ band hybridization (proportional to the curvature of graphene flake). Other effects of electron-flexural phonons coupling in graphene have been  disscused in the literature \cite{Ocha_2012}. Also, we need to take into account interactions with the substrate. Let us write first the contribution from the pseudoelectromagnetic fields, this is given by \cite{Morpurgo_2006,Guinea2009,Rainis_Guinea_2011},}

\begin{equation} \label{eq:Dirac-type Hamiltonian}
\mathcal{\hat{H}}_{\eta}(\boldsymbol{r})=v_{F} \boldsymbol{\sigma}_{\eta} \cdot \left( \boldsymbol{\hat{p}}- \eta \boldsymbol{A}(\boldsymbol{r},t) \right)+V(\boldsymbol{r},t) \sigma_0,
\end{equation}

where $\boldsymbol{r}=(x,y)$ is the position vector, the subscript $\eta=\pm 1$ labels the Dirac points $\boldsymbol{K},\boldsymbol{K}\,'$ respectively; $v_{F}$ is the Fermi velocity ($v_{F}/c \approx 1/300$ with $c$ is the vacuum speed of light); $\boldsymbol{\hat{p}}=(\hat{p}_{x}, \hat{p}_{y})$ is the moment operator for the charge carriers, $\boldsymbol{\sigma}=(\eta \sigma_{x}, \sigma_{y})$ is the Pauli matrix vector and $\sigma_0$ the $2\times2$ identity matrix, and $\boldsymbol{A}$ and $V$ are the pseudo vector and scalar potentials respectively, given by \cite{Guinea2009,Naumis2017,Oliva2015,Bastos2018}

\begin{eqnarray} 
V(\boldsymbol{r},t)=g(\varepsilon_{xx}+ \varepsilon_{yy}) \label{eq:general scalar potential}\\
\boldsymbol{A}(\boldsymbol{r},t)=(A_{x},A_{y})
= \frac{\hbar \beta}{2 a_{cc}}(\varepsilon_{xx}-\varepsilon_{yy},-2 \varepsilon_{xy}) \label{eq:general vector potential}
\end{eqnarray}
The parameter  $a_{cc}=1{.}42$ \r{A} is the interatomic distance for undeformed  graphene lattice and the dimensionless coefficient  $\beta \approx 3{.}0$ measures the effect of the deformation on the hopping parameter. {\color{black}The coupling $g$ was thought first to be around\cite{Vozmediano2010} $20$ eV , however, this turned out to be a bare estimation as charge screening leads to a much lower renormalized value \cite{katsnelson_2020} $g  \approx 4$ eV.}
The coefficient $g$ refers to flexural changes in the membrane while the term  ${\hbar \beta }/{2a_{cc}}$ refers to changes in bond length, as we know it requires more energy to make bond length changes to a rearrangement in the positions of the atoms on the membrane \cite{Suzuura2002}.{\color{black} Therefore, depending on the deformation, the value of  $g$ is within a range of energies while the factor ${v_F\hbar \beta}/{2a_{cc}}\approx{6.932}$ eV remains approximately constant (within the validity range of our model).} 

In general, we can consider a displacement outside the plane $h=h(\boldsymbol{r},t)$, and a displacement inside the plane  $\boldsymbol{u}=\boldsymbol{u}(\boldsymbol{r},t)$. The stress tensor $\varepsilon_{\mu \nu}$ is given by
\begin{equation} \label{eq:general stress tensor}
\varepsilon_{\mu \nu}= \frac{1}{2} \left( \partial_{\mu} h \partial_{\nu} h \right)+ \frac{1}{2} \left( \partial_{\mu} u_{\nu}+ \partial_{\nu} u_{\mu} \right), \,\,\, \mu, \nu =x,y.
\end{equation}
We shall consider the simplest case, in which the deformation is only perpendicular to the plane, i.e., $\boldsymbol{u}=0$, so from Eq. \eqref{eq:general stress tensor} 
\begin{equation} \label{eq:particular stress tensor}
\begin{split}
\varepsilon_{xx}&= \frac{1}{2} \left( \partial_{x} h \right)^{2}\\
\varepsilon_{yy}&= \frac{1}{2} \left( \partial_{y} h \right)^{2}\\
\varepsilon_{xy}&= \frac{1}{2} \left( \partial_{x} 
h\right) \left( \partial_{y} h \right)
\end{split}
\end{equation}

{\color{black}  We now discuss whether to include or not corrections to Eq.  (\ref{eq:Dirac-type Hamiltonian}) depending on the experimental scenario. Out of plane hybridize $\pi$ orbitals with higher
orbitals of carbon, leading to a first-order contribution in the
spin-orbit interaction strength, contrary to in-plane distortions,
whose contribution is at least quadratic \cite{Ocha_2012}. The corrections to the Hamiltonian are given by \cite{Ocha_2012},
\begin{equation}
    \mathcal{\hat{H}}=\mathcal{\hat{H}}_{A_1}+\mathcal{\hat{H}}_{B_2}+\mathcal{\hat{H}}_{G'}
\end{equation}
where the labels $A_1,B_2,G'$ are the irreducible representations of the group $C_6^{"}$, resulting from considering a graphene's unit cell with six atoms, used in such a way to avoid dealing with degenerate states at two inequivalent Dirac points \cite{Ocha_2012}. Such corrections leads to a Kane-Mele mass and a Rashba-like coupling present only in the case of a mirror symmetry breaking.  Both coupling effects are weak \cite{Morpurgo_2006,Ocha_2012}, as the estimates are in the range of $1-15$ $\mu $eV. 
, for the present work such effects can be safely neglected as a first approximation. 

Also,  the $\sigma-\pi$ orbitals hybridization leads to a correction to $V(\bm{r})$ as we need to add in the diagonal of the Dirac equation the following potential, \cite{Kim_2008,Rainis_Guinea_2011},
\begin{equation}
    V_{\pi\sigma}(\bm{r})=-g_1(\nabla ^{2} h)^{2}
\end{equation}
where $g_1=3\alpha/4a_{cc}$ and $\alpha \approx 9.23$ eV. The resulting $ V_{\pi\sigma}(\bm{r})$ from local curvature will
off-set the charge neutrality point from the average chemical potential  \cite{Kim_2008}. 

In the same line of reasoning, a substrate flexural deformations would be accompanied by the variation of on-site energies of carbon orbitals. This can be treated by decomposing the interaction into a smooth spatial effective potential\cite{Rainis_Guinea_2011} $V_{sub}(\bm{r})\sigma_0$ and, if the substrate is such that produces a bipartite symmetry breaking, an extra term\cite{Morpurgo_2006} $\Delta(\bm{r})\sigma_z=V_A(\bm{r})-V_B(\bm{r})\sigma_z$, which measures the difference of the electrostatic potential in the two sublattices $A$ and $B$, for example, due to charges located at random position in the substrate supporting graphene.
The inclusion of this term depends upon the kind of substrate, for example, in SiO$_2$ such component can be neglected as graphene  follows the substrate
potential in a coarse-grained and smooth manner \cite{Morpurgo_2006} or in graphene over oxidized Cu (111) surface \cite{Gottardi2015}. For simplicity, here we will consider substrates in which the local potential $\Delta(\bm{r})\sigma_z$ can be neglected. Therefore, the hybridization and substrate effects can be taken into account by making the following replacement in Eq. (\ref{eq:Dirac-type Hamiltonian}),
\begin{equation}
    V(\bm{r},t) \rightarrow V(\bm{r},t)+V_{sub}(\bm{r},t)+V_{\pi\sigma}(\bm{r},t)
\end{equation}
For certain substrates as oxidized Cu (111) surface, a high-k dielectric material, the alteration of graphene due to electrostatic effects is minimal and in fact $V_{sub}(\bm{r},t)$ can be neglected \cite{Gottardi2015}. 
}

To simplify the resulting equations, we introduce new variables defined as,
\begin{equation} \label{eq:random variables}
\begin{split}
 l_{1}(\boldsymbol{r},t) &\equiv  \left( \partial_{x} h \right)^{2} -\left( \partial_{y} h \right)^{2}\\
 l_{2}(\boldsymbol{r},t)& \equiv 2 \left( \partial_{x} h \right) \left( \partial_{y} h \right)
\end{split}
\end{equation}
which will give us information about how ``strong"{} the vertical displacements are. On the other hand, by making use of the Eqs. \eqref{eq:general scalar potential}, \eqref{eq:general vector potential},\eqref{eq:particular stress tensor} and \eqref{eq:random variables}, we can rewrite the scalar and pseudo-vector potentials,
\begin{equation} \label{eq: particular potentials}
\begin{split}
V(\boldsymbol{r},t)&= \frac{g}{2} \sqrt{l_{1}^{2}(\boldsymbol{r},t)+l_{2}^{2}(\boldsymbol{r},t)}\\
\boldsymbol{A}(\boldsymbol{r},t)&= \frac{\hbar \beta}{4 a_{cc}} \left[ l_{1}(\boldsymbol{r},t) \hat{x}- l_{2}(\boldsymbol{r},t) \hat{y} \right].
\end{split}
\end{equation}
{\color{black} From Eq. \eqref{eq:Dirac-type Hamiltonian} and \eqref{eq: particular potentials},  the Hamiltonian is
\begin{equation} \label{eq:Hamiltonian 1}
\boldsymbol{\mathcal{\hat{H}}}_{\eta}(\boldsymbol{r})=\boldsymbol{\mathcal{\hat{H}}}_{0}(\boldsymbol{r})+\boldsymbol{W}(\boldsymbol{r},t)+  V_{eff}(\boldsymbol{r},t)\sigma_0    
\end{equation}
with,
\begin{equation}
    V_{eff}(\boldsymbol{r},t)=\left(\frac{g}{2} |l(\boldsymbol{r},t)|-g_1 (\nabla^{2}h)^{2} \right)
\end{equation}
 The hat is used to denote the differential operators},
\begin{equation} \label{eq:Hamiltonian}
\begin{split}
\boldsymbol{\mathcal{\hat{H}}}_{0}(\boldsymbol{r})&= v_{F} \left( \begin{array}{lcc}
0 & (\eta \hat{p}_{x}-i \hat{p}_{y})\\
\eta \hat{p}_{x}+ i \hat{p}_{y} & 0 
\end{array} \right)\\
\boldsymbol{W}(\boldsymbol{r},t)&= \left( \begin{array}{lcc}
0 & - \eta \tilde{\beta} l(\boldsymbol{r},t)\\
- \eta \tilde{\beta} l^{*}(\boldsymbol{r},t)& 0
\end{array} \right)
\end{split}
\end{equation}
where $l(\boldsymbol{r},t) \equiv   \eta l_{1}(\boldsymbol{r},t)+ i l_{2}(\boldsymbol{r},t) $ and we defined the parameter $\bar{\beta}$ as,
\begin{equation}
    \bar{\beta}= \frac{v_F\hbar \beta}{4a_{cc}} \approx 3.476 \text{ eV}
\end{equation}

The dynamic equation for the spinor $\Psi_{\eta}(\boldsymbol{r},t)$ follows a time-dependent Schrödinger type equation
\begin{equation} \label{eq:time-dependent schrodinger equation}
\begin{split}
i \hbar \frac{\partial}{\partial t} \Psi_{\eta}(\boldsymbol{r},t)= \boldsymbol{\mathcal{\hat{H}}}_{\eta}(\boldsymbol{r}) \Psi_{\eta}(\boldsymbol{r},t)
\end{split}
\end{equation}
where 
\begin{equation} \label{eq:spinor vector form }
\Psi_{\eta}(\boldsymbol{r},t)= \left( \begin{array}{cc}
     \psi_{A}^{\eta}(\boldsymbol{r},t) \\
     \psi_{B}^{\eta}(\boldsymbol{r},t)
\end{array}\right)
\end{equation}
 It is straightforward to prove that the Schrödinger type equation \eqref{eq:time-dependent schrodinger equation} can be rewritten as
\begin{equation} \label{eq:time-dependent schrodinger equation for both spinor components}
\begin{split}
    i \hbar \frac{\partial \psi_{A}^{\eta}}{\partial t}&= V_{eff}(\boldsymbol{r},t) \psi_{A}^{\eta}+ \left[v_{F}(\eta \hat{p}_{x}-i \hat{p}_{y})- \eta \bar{\beta} l \right] \psi_{B}^{\eta},\\
    i \hbar \frac{\partial \psi_{B}^{\eta}}{\partial t}&= V_{eff}(\boldsymbol{r},t) \psi_{B}^{\eta}+ \left[v_{F}(\eta \hat{p}_{x}+i \hat{p}_{y})- \eta \bar{\beta} {l^{*}} \right] \psi_{A}^{\eta}.\\
\end{split}    
\end{equation}

Notice how the magnitude of the disorder enters in the Dirac equation through the parameters $\bar{\beta}$ and $g$. While
$g$ plays the role of a random local chemical potential,
$\tilde{\beta}$ is a random local magnetic field.

Eq.  (\ref{eq:time-dependent schrodinger equation for both spinor components}) is a complex stochastic equation. Instead of solving the time-dependent problem, we consider that the deformation process is adiabatic in the time scale of the electron dynamics. In such a case, we can suppose that
the disorder is quenched and thus $l_1$ and $l_2$  are time-independent. In such a case,  Eq. (\ref{eq:Hamiltonian})
becomes a time-independent Hamiltonian with a spatial random potential  $l(\boldsymbol{r},t)=l(\boldsymbol{r})$.  This is the case of topographic corrugations, such as wrinkles and foldings \cite{Wei2018,Verhagen2019}.

Returning to Eq. \eqref{eq:time-dependent schrodinger equation} becomes the time-independent Schrödinger equation $ \boldsymbol{\mathcal{\hat{H}}}_{\eta}(\boldsymbol{r}) \Psi_{\eta}(\boldsymbol{r})= E \Psi_{\eta}(\boldsymbol{r})$ and we are interested in finding the distribution of the Hamiltonian eigenvalues of $\boldsymbol{\mathcal{\hat{H}}}_{\eta}(\boldsymbol{r})$ and the wavefunctions.

\section{Folded  Deformations} \label{sec:Folded deformations}

To understand the changes induced by random
flexural deformations, we study 
folded deformations. Such kind of fields have been
observed experimentally in deformed graphene \cite{Kim2011,YI2016,Hallam2015}
and there are some studies for particular
deformations \cite{RCarrillo2016,Nancy2018,Sandler2018}.
In a general folded deformation, the field does
not vary in one direction. Therefore, it can be written as,

\begin{equation}
    h(y)=\sum_{k=-k_c}^{k_c} a_k \exp(iky) 
\end{equation}
 with $a_{-{k}}=a_{{k}}^{*}$ as $h({y})$ is a real,  and the coefficients $a_{{k}}$ can be deterministic or random variables. $k_c$ is a cutoff parameter and in what follows all sums are understood to use it. {\color{black} $k_c$ can be estimated from the Bose-Einstein distribution and depends upon the experimental conditions (see Appendix \ref{sec:Cutoff}). }
 
 From Eqs. (\ref{eq:general vector potential}) and  (\ref{eq:general stress tensor}) , 
 the vectorial potential has only one component different from zero,
 \begin{equation}
     A_x(y)=\frac{\hbar \beta}{4 a_{cc}}
    \left[ \sum_{k} a_k k\exp(iky)  \right]^{2}
 \end{equation}

 The advantage of this particular deformation
is that $\bm A(\bm{r})$ is in the Coulomb gauge, as it satisfies $\nabla \cdot \bm A(\bm{r})=0$, therefore can be obtained as the derivative of a scalar field, 

\begin{equation}
    A_i=\epsilon_{ij}\partial_j \Phi(\boldsymbol{r}) 
    \label{eq:DefPotential}
\end{equation}
where $\epsilon_{ij}$ is the 2D Levi-Civita tensor with $i=x,y$ and $j=x,y$. For this particular case,  we express $\Phi(y)$ in terms of the following Fourier decomposition, 
 \begin{equation}
     \Phi(y)=\Phi_0(y)+
     \sum_{k \ne 0}e^{iky} \tilde{\Phi}(k)\label{eq:Phiy}
 \end{equation}
with,
\begin{equation}
    \Phi_0(y)=\frac{\hbar \beta}{4 a_{cc}}\left(\sum_k k^{2}|a_k|^{2}\right)y
\end{equation}
and,
\begin{equation}
     \tilde{\Phi}(k)=-i\frac{\hbar \beta}{4 a_{cc}k} \left[
     \sum_{k'}a_{k}a_{k'-k}^{*}k'\left(k'-k\right) \right]\label{eq:FourierPhiy}
 \end{equation}

The associated pseudomagnetic field is $\bm{B}=\bm{\nabla}^{2} \Phi (\bm{r})$. It is worthwhile noticing that although $\Phi_0(y)$ does not produce a pseudomagnetic field, it produces an  Aharonov-Bohm like effect as it leads to a constant $\bm A(\bm{r})$.  {\color{black} Finally, the contribution from the $\sigma_0$ term is,
\begin{equation}
    V_{eff}(y)=\frac{g}{2}|l_1(y)|-\frac{g_1}{4|l_1(y)|}\left(\frac{\partial |l_1(y)|}{\partial y}\right)^{2}
\end{equation}
with $l_1(y)=v_FA_x(y)/\bar{\beta}$.} 

\begin{figure}[h!]
\includegraphics[height=5.7cm, width=8.7cm]{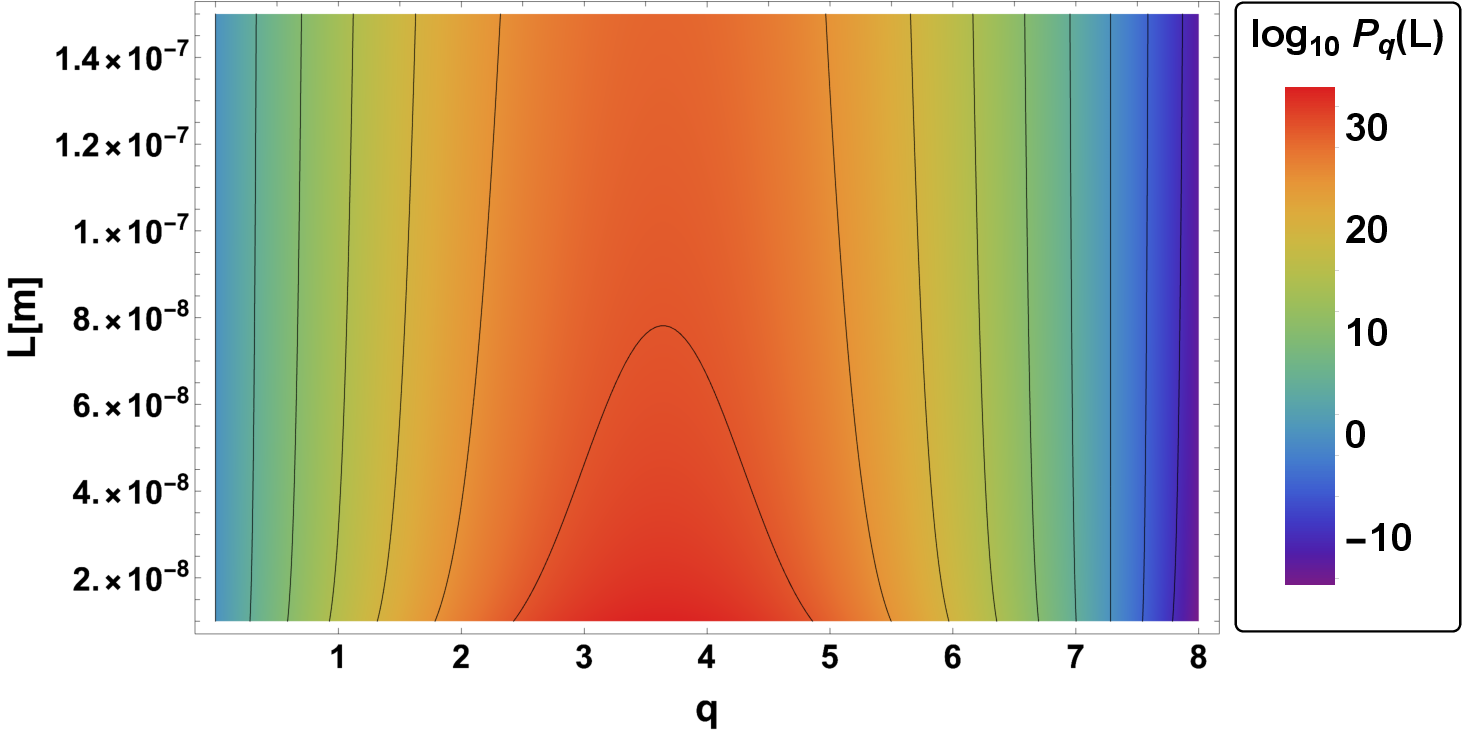}
\caption{{\color{black}Multifractality of zero modes wave functions. Contour plot of $P_q(L)$ as a function of the sample length $L$ and the exponent $q$, for $\Delta_A=1$, chosen to be below the quantum phase transition to the Hall effect at $\Delta_A=\pi$.}}.
\label{fig:DOS}
\end{figure}

\begin{figure}[h!]
\includegraphics[scale=0.41]{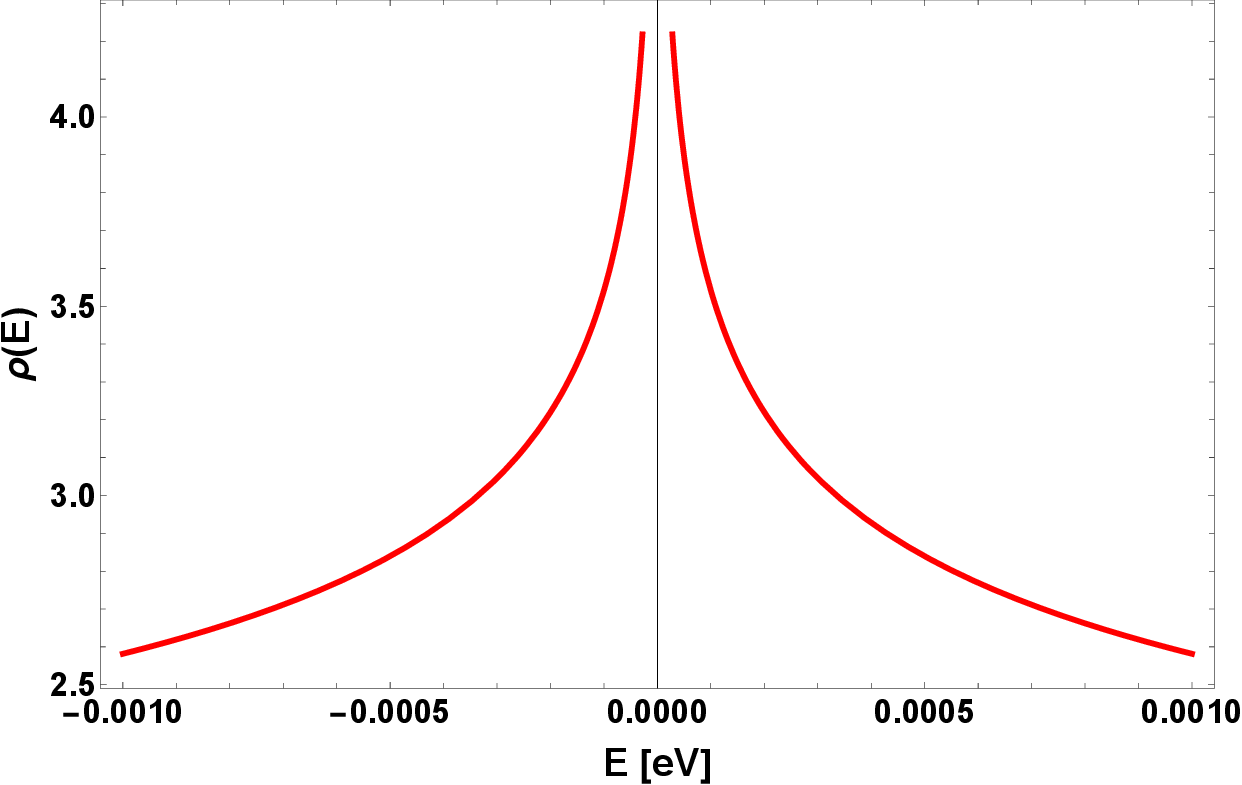}
\caption{{\color{black}Density of states (DOS) as a function of the energy and around the Dirac point with added random pseudomagnetic fields, with $\Delta_A=1$, chosen to be below the quantum phase transition to the Hall effect at $\Delta_A=\pi$.}}.
\label{fig:Pvsq}
\end{figure}

An interesting consequence of having a field derived from the potential is that for any flexural field, being deterministic or random, the zero-mode can always be constructed. {\color{black} Zero modes in the Dirac equation are topologically protected thus their existance is independent of $V_{eff}(y)$. As a consequence, the usual approach is to neglect such contribution keeping only the pseudomagnetic field  \cite{katsnelson_2020}. Also, the contribution from $g$ and $g_1$ tends to cancel each}. Therefore, from the Schrödinger and Eq. (\ref{eq:DefPotential}), we obtain that for $E=0$ the wave function is,

\begin{equation}
    \psi_{\pm}(\bm{r})=(const.) (1\pm \sigma_z)
  \left(  \begin{array}{c}
        e^{\Phi(y)}   \\
        e^{-\Phi(y)}  
    \end{array} \right)
\end{equation}
where $\sigma_z$ is the Pauli $z$ matrix.
Similar functions were studied years ago in the context of the integer quantum Hall transition \cite{Andreas1994}. It can be proved that for a random magnetic field in which the vector potential 
satisfies a Gaussian white-noise distribution
with mean zero and variance $\Delta_{A}$ such that the average coefficients in Eq. (\ref{eq:Phiy}) are,
\begin{equation}
    \langle \tilde{\Phi}(k)\tilde{\Phi}(k')\rangle=(2\pi)^{2}\delta(k-k')\frac{\Delta_A}{k^{2}}
\end{equation}
while the resulting wave-function is multifractal \cite{Andreas1994}. In a  sample of size $L \times L$, the moments of the participation ratio $P_q(L)$ that measures a multifractal localization \cite{Barrios_Vargas_2012},
\begin{equation}
    P_q(L)=\langle |\psi({\bm{r})}|^{2q}\rangle
\end{equation}
are given by  \cite{Andreas1994},
\begin{equation}
     P_q(L) \approx \frac{1}{L^{2+\tau(q)}}
\end{equation}
with, 
\begin{equation}
    \tau(q)=2(q-1)+\frac{\Delta_A}{\pi}q(1-q)
\end{equation}
where $q$ need not be integer. In Fig. \ref{fig:Pvsq} we present a surface plot of $P_q(L)$ for a $\Delta_A$ below the quantum phase transition that occurs at $\Delta_A=\pi$.
For big samples, the multifractal spectrum is dominated by its maximal value, from where the typical participation is  \cite{Andreas1994}, 

\begin{equation}
    P_{\text{typical}}(L)=e^{\langle \ln|\Psi|^{2} \rangle}\approx \frac{1}{L^{2+\Delta_A/\pi}}
\end{equation}
Around these states and near the Fermi energy, the density of states (DOS) is \cite{Andreas1994},

\begin{equation}
    \rho(E)=E^{\frac{2-z}{z}}
\end{equation}
with $z=2+\Delta_A/\pi$. Fig. \ref{fig:DOS} presents the resulting DOS showing that the main effect is an incresead density at the Dirac point. The wavefunction multifractality and the power law DOS means that an unusual electron velocity distribution will appear even in the simplest case of a Gaussian random flexural field, without restoring to Levy distributions of membrane jumps in graphene. In any case, the Levy jumps will induce an even more unusual distribution.

We end up by considering the particular contribution of the Aharonov-Bohm term which  for some geometries produces interesting effects in graphene \cite{deJuan2011}, nevertheless has not been studied for random fields. First we write the Fourier coefficients $a_k$ as the sum of an average plus a fluctuation part, $a_k= \langle a_k \rangle +\delta a_k$. If $a_k$ is Gaussian distributed with zero mean we have,
\begin{equation}
    \Phi_0(y)=\frac{\hbar \beta}{4 a_{cc}}\sum_k (\delta a_{k})^{2}k^{2}y \approx  \frac{\pi}{6} \frac{\hbar \beta}{ a_{cc}}\Delta_A k_c^{3}y
\end{equation}
and thus the phase difference between particles, with the same start and end points, but travelling along two different paths is,
\begin{equation}
    \Delta \mathcal{\phi} =\left(\frac{d\Phi_0(y)}{dy}\mathcal{A} \right)\frac{e}{\hbar}=\frac{\pi}{6}\frac{\beta}{ a_{cc}} (\Delta_A k_c^{3}\mathcal{A}) e
    \label{eq:phasediference}
\end{equation}
where $\mathcal{A}$ is the area bounded by the two paths as seen in Fig. \ref{fig:Ripple}. For thermally activated fields, $k_c$ is determined from the temperature ($T$) population given by the Bose-Einstein distribution. As $\Delta_A \sim k_BT$, Eq. (\ref{eq:phasediference})  implies a very strong  temperature dependent phase shift. This result is in agreement with recent first-principles calculations based on density functional theory and the Boltzmann equation\cite{Tue_2020}.

\section{Conclusions. \label{sec: conclusions}}

We studied the effects  in the electronic properties of graphene of folded flexural deformations, which are equivalent to  electromagnetic fields in the Columb gauge.  First we studied general  folded deformations giving an expression for the zero-modes which are the ones at the Fermi level for half-filled systems. For random Gaussian distributed folded deformations, we made contact with works on the quantum Hall effect under random magnetic fields, showing that the wave functions are multifractal and the density of states has a power law behavior. This indicates that the system can present interesting behaviors. In particular,  there is a remarkable Aharonov-Bohm pseudo-effect. {\color{black}The wavefunction multifractality can be observed as an unusual dependence of the conductance with the length or by an unusual electron velocity distribution, as has been observed in some experiments \cite{Ackerman2016}.
In fact, there are clear signatures of such zero modes
exotic multifractal conductance fluctuations  in recent experiments with high-mobility single-layer graphene field-effect transistors \cite{Amin2018}}.

\begin{acknowledgments}
We are grateful to Alejandro Pérez Riascos  for their support and feedback in carrying out this work. We thank UNAM-DGAPA PAPIIT project IN102620 and CONACYT project 1564464. A.E.C. thanks CONACYT for providing a schoolarship.
\end{acknowledgments}
\section*{Data Availability}
The data that support the findings of this study are available
from the corresponding author upon reasonable request.
\appendix

\section{Cut-off criteria for the deformation field}\label{sec:Cutoff}

{\color{black}Consider the displacement outside the plane, in general it can be written as,
 \begin{equation} \label{eq: displacement function in RPA }
    h(\bm{r})=\sum_{\boldsymbol{k=-k_c}}^{kc}  a_{\bm{k}} \exp(i\bm{kr})
\end{equation}

It is important to remark that for graphene, the room temperature is far below the Debye temperature \cite{Feng2018,Balandin2020}, which is about $1000$ and $2300$ K , and therefore $\hbar \omega(k_c) \leq k_{B}T $. As the purely harmonic flexural dispersion goes as \cite{Castro_2010} $\omega(\bm{k})=\alpha |\bm{k}|^{2}$ with $\alpha=4.6 \time 10^{2} m^{2}/s$, it follows that,
 \begin{equation}
 |\bm{k}_c| \approx \sqrt{\frac{k_{B}T}{\alpha \hbar}}   
 \end{equation}
Notice that $k_c$ is a strain dependent quantity \cite{Castro_2010,Zakharchenko_2010}. In fact, for free-standing graphene at thermal equilibrium, 
 \begin{equation} \label{eq: flexural field}
     \left \langle a_{\boldsymbol{k}}^{2}\right\rangle= \frac{\hbar (1+2n_{B}(\omega(\boldsymbol{k})))}{2M_{C} \omega(\boldsymbol{k})} \approx \frac{k_{B} T}{M_{C} \omega(\boldsymbol{k})}
 \end{equation}
 where  $n_{B}(\omega(\boldsymbol{k}))$ is the thermal population of mode $\boldsymbol{k}$, $M_{C}$ is the carbon mass and the second equality holds when $\hbar \omega(\boldsymbol{k}) \ll k_{B}T$. For purely harmonic flexural modes the fluctuation $\langle a_{\boldsymbol{k}}^{2} \rangle$ diverges as $|\boldsymbol{k}|^{-4}$ for small $\boldsymbol{k}$. In real samples, however, it is known that the singularity gets renormalized
 due to lattice imperfections (i.e., by anharmonic effects). The resulting dispersion can be parametrized as  $\omega(k)= \alpha \sqrt{k^{4}+k^{4-\tau}k_{c}^{\tau}}$ for $\tau >0$, from where it follows that the quadratic mean displacement of each field
 mode, in the long wavelenght, is given by \cite{Fratini2013},
 \begin{equation}
     \langle |a_k|^{2} \rangle \propto \frac {k_{B}T}{k^{4-\tau}k_c^{\tau}}
 \end{equation}
 where $\tau$ depend on the physical mechanism of renormalization.
 The physical scenarios are \cite{Fratini2013} : a) substrate pinning that opens a gap in the phonon spectrum, corresponding to $\tau=4$; b) strain which makes the dispersion linear at long wavelengths,  $\tau=2$, c) anharmonic effects which yield $\tau=0{.}82$. }

\section{Coulomb norm for general pseudo-electromagnetic fields }\label{sec:Coulomb}

Although some works assume the Coulomb norm for general pseudo-electromagnetic fields \cite{Kailasvuori_2009}, let us show that in general such deformation can not be written as the derivative of a scalar field. This can be proved as follows, if we consider $A_{i}(\boldsymbol{r})= \epsilon_{ij} \partial_{j} \Phi(\boldsymbol{r})$ with  $\Phi(\boldsymbol{r})= \sum_{\boldsymbol{k}} b_{\boldsymbol{k}} \exp(i \boldsymbol{k} \cdot \boldsymbol{r})$ it holds that
\begin{equation} \label{eq:condition on b_k}
    \begin{split}
        b_{\boldsymbol{k}}&= - \frac{i v_{F} \tilde{\beta}}{k_{y}} \sum_{\boldsymbol{k}'} a_{\boldsymbol{k}} a_{\boldsymbol{k}'}^{*} (k_{x}k_{x}'-k_{y}k_{y}') e^{-i \boldsymbol{k}' \cdot \boldsymbol{r}}\\
        \text{and }  b_{\boldsymbol{k}}&= - \frac{i v_{F} \tilde{\beta}}{k_{x}} \sum_{\boldsymbol{k}'} a_{\boldsymbol{k}} a_{\boldsymbol{k}'}^{*} (k_{x}k_{y}'+k_{y}k_{x}') e^{-i \boldsymbol{k}' \cdot \boldsymbol{r}}.
    \end{split}
\end{equation}
In general, the system of equations in Eq. \eqref{eq:condition on b_k} has no solutions for $b_{\boldsymbol{k}}$ except for few particular cases, as the folded potential studied here.

\section*{References}
\bibliographystyle{unsrt}
\providecommand{\noopsort}[1]{}\providecommand{\singleletter}[1]{#1}%



\end{document}